\documentclass[superscriptaddress,twocolumn,showpacs,preprintnumbers,amsmath,amssymb,pra]{revtex4}
\usepackage{epsfig}
\usepackage{dcolumn}
\newcommand{\bm}[1]{ \mbox{\boldmath $#1$}  }
\newcommand{\ud}{\mathrm{d}}
\begin{document}

\title{Thomas-Fermi Approximation for a Condensate with Higher-order Interactions}

\author{M. Th{\o}gersen}
\affiliation{Department of Physics and Astronomy, University of
  Aarhus, DK-8000 Aarhus, Denmark}

\author{N.~T. Zinner}
\affiliation{Department of Physics, Harvard University, Cambridge,
  Massachusetts 02138, USA}

\author{A.~S. Jensen}
\affiliation{Department of Physics and Astronomy, University of
  Aarhus, DK-8000 Aarhus, Denmark}

\date{\today}

\hyphenation{Fesh-bach}

\begin{abstract}
We consider the ground state of a harmonically trapped Bose-Einstein condensate within the Gross-Pitaevskii theory
including the effective-range corrections for a two-body zero-range potential. The 
resulting non-linear Schr{\"o}dinger equation is solved analytically in the Thomas-Fermi approximation neglecting
the kinetic energy term. We present results for the chemical potential and the condensate
profiles, discuss boundary conditions, and compare to the usual Thomas-Fermi approach. We discuss 
several ways to increase the influence of effective-range corrections in experiment with magnetically 
tunable interactions. The level of tuning required could be inside experimental reach in the near future.
\end{abstract}
\maketitle

\section{INTRODUCTION}
The Gross-Pitaevskii (GP) equation
\cite{dalfovo99,pethick02,pitaevskii03} has been extremely
successful in describing a wide range of mean-field features for
experiments with Bose-Einstein condensates (BECs).  In particular, the
Thomas-Fermi (TF) approximation
\cite{baym96,dalfovo96,lundh97}, where the kinetic
energy is neglected, has been very rewarding \cite{hau98}. This
approximation holds for repulsive condensates with positive scattering
length $a$ and large particle numbers.
In the regime of
validity of the TF approximation, the total energy is distributed between
interaction energy and potential energy from the confining trap, while
the kinetic energy becomes negligible.

Because of the non-linear nature of the GP equation, it is only solved
analytically in a few cases, e.g., vortices
and solitons in homogeneous condensates \cite{pethick02,pitaevskii03}. 
The TF solution is also analytical, although it only holds in
the bulk of the condensate. At the surface the approximation breaks
down and is usually patched by including the kinetic energy at the surface 
\cite{dalfovo96,lundh97}.

The interactions of the ordinary GP equation are based on the lowest
order zero-range potential, which is governed by the scattering length
alone. Although this approximation is usually very good, the
higher-order corrections to the scattering dynamics
\cite{roth01,fu03,collin07} can be crucial in certain cases, e.g.,  for
Rydberg molecules embedded in BECs \cite{collin07} and for narrow
Feshbach resonances \cite{zinner2009}. Inclusion of higher-order terms 
is well known and applied in
Skyrme-Hartree-Fock calculations in nuclear physics \cite{brack85}. Here, they
often play a crucial role in order to get bulk nuclear properties right \cite{sky56,sie87}.
However, the effects of similar higher-order terms in the GP equation
have been less investigated.

In this paper, we solve the modified GP equation with higher-order
interactions analytically in the TF approximation. The paper is
organized as follows. In Sec.~\ref{MGPeq}, we introduce the 
modified GP equation and its parameters and show how it is derived from 
an appropriate energy density functional with careful treatment of
boundary terms. We present the analytical solution in the TF approximation
in Sec.~\ref{TFA} and discuss the condensate size and chemical potential 
as function of the interaction parameters in Sec.~\ref{sizechem}. The
density profiles and energies are discussed in Sec.~\ref{densene}, and 
in Sec.~\ref{cons}, we address the consistency of the TF approximation by
considering the kinetic energy of the solutions. We compare to some relevant
atomic systems in Sec.~\ref{compare} and finally 
present our conclusions in Sec.~\ref{conc}.

\section{MODIFIED GP EQUATION}\label{MGPeq}
We assume that the condensate can be described by the GP equation.
Since we are interested in the ultracold regime, where the temperature
is much smaller than the critical temperature for condensation, we
adopt the \mbox{$T=0$} formalism. In order to include higher-order
effects in the two-body scattering dynamics, we use the modified GP
equation derived in \cite{collin07}, which in the stationary form reads
\begin{equation}
  \mu\Psi
  =\left[-\frac{\hbar^2}{2m}\nabla^2
    +V(r)+U_0\left(|\Psi|^2+g_2\nabla^2|\Psi|^2\right)
  \right]\Psi,
\label{GPE}
\end{equation}
where $m$ is the atomic mass, $V$ is the external trap,
$U_0=4\pi\hbar^2 a/m$, and $g_2=a^2/3-ar_e/2$, with $a$ and $r_e$
being, respectively, the $s$-wave scattering length and effective range
\cite{collin07}. We assume an isotropic trap, $V(r)=m\omega^2 r^2/2$, and introduce the
trap length $b=\sqrt{\hbar/m\omega}$. The single-particle density,
$\rho(r)=|\Psi(r)|^2$, is normalized to the particle number, $N=\int\ud{\bm r}\rho(r)$,
and $\mu$ is the chemical potential.

As the boundary conditions are important for the TF 
approximation applied below we now discuss the procedure for obtaining
the modified GP equation from the corresponding energy functional which is
\begin{equation}\label{efunc}
  E(\Psi)=\int \ud{\bm r} (
  \epsilon_K +\epsilon_V +\epsilon_I +\epsilon_{I2} ),
\end{equation}
with kinetic, potential, and interaction energy densities 
\begin{eqnarray}
  &\epsilon_K=\frac{\hbar^2}{2m}|\nabla\Psi|^2,\qquad 
  &\epsilon_V=V(\bm r)|\Psi|^2,\\
  &\epsilon_I=\frac{1}{2}U_0 |\Psi|^4,\qquad
  &\epsilon_{I2}=\frac{1}{2}U_0 g_2|\Psi|^2 \nabla^2|\Psi|^2.
\end{eqnarray}
The corresponding integrated energy contributions are denoted $E_K$, $E_V$,
$E_I$, and $E_{I2}$, respectively. 
To obtain Eq.~\eqref{GPE}, we vary Eq.~\eqref{efunc} with respect to $\Psi^*$ for fixed $\Psi$. 
To first
order in $\delta\Psi^*$, we have
\begin{equation}\label{GPvar}
  \begin{split}
    \delta E
    &=E[\Psi^*+\delta\Psi^*]-E[\Psi^*]\\
    &=\int\ud{\bm r} \bigg[ -\frac{\hbar^2}{2m}\nabla^2\Psi+ V(\bm r)\Psi\\
    &\qquad\qquad+U_0\left( |\Psi|^2+g_2\nabla^2|\Psi|^2\right)\Psi \bigg] \delta\Psi^*\\
    &+\int\ud{\bm S}\cdot |\Psi|^2\nabla\left(\Psi\delta\Psi^*\right)
    -\int\ud{\bm S}\cdot \Psi\delta\Psi^*\nabla|\Psi|^2\\
    &+\int\ud{\bm S}\cdot \delta\Psi^*\nabla\Psi.
  \end{split}
\end{equation}
Here, $\bm S$ is the outward-pointing surface normal.  In the usual
analysis, one assumes that $\Psi$ and $\nabla\Psi$ vanishes at
infinity, drops the boundary terms, and Eq.~\eqref{GPE} is obtained by
varying $E-\mu N$.  However, the existence of these surface terms is
essential for the inclusion of higher-order interactions as discussed below.

In the rest of this paper we use trap units, $\hbar\omega=b=1$,  i.e.,
energies ($E$, $V$, $\mu$, etc.) are measured in units of $\hbar\omega$ and
lengths ($a$, $r_e$, $r$, etc.) in units of $b$. Note that 
$g_2$ has dimension of length squared.

\section{THOMAS-FERMI APPROXIMATION}\label{TFA}
Let us briefly review the standard Thomas-Fermi approximation
\cite{baym96,dalfovo99,pethick02,pitaevskii03}. Neglecting the kinetic-energy 
term, as compared to the trap and interaction energies, the GP
equation has the solution
\begin{equation}
  \label{eq:density-TF}
  \rho_{TF}=\frac{1}{4\pi a}(\mu_{TF}-\frac{1}{2}r^2),
\end{equation}
with chemical potential $\mu_{TF}$. This solution is used out to the
surface, $R_{TF}$, while outside $\rho_{TF}=0$. The
normalization and surface condition $\rho_{TF}(R_{TF})=0$ give
\begin{equation}
  \label{eq:mu-rmax-TF}
  \mu_{TF}=\frac{1}{2}R_{TF}^2.\qquad
  R_{TF}=(15Na)^{1/5}.
\end{equation}
The total energy becomes
\begin{equation}
  \frac{E_{TF}}{N}=\frac{5}{7}\frac{R_{TF}^2}{2}.
\end{equation}
The trap and interaction energies are $E_V=3E/5$ and $E_I=2E/5$,
respectively. Since $R_{TF}>0$ in Eq.~\eqref{eq:mu-rmax-TF}, these
results only hold for $a>0$.
The TF approximation is good for $Na\gg 1$, except at the surface
region where the kinetic-energy density diverges. Here, the solution
can be corrected as in \cite{dalfovo96,lundh97,pethick02,pitaevskii03}, essentially
giving a small exponential tail.

\subsection*{Inclusion of higher-order interactions}
We now consider the TF approximation with the
higher-order interaction term, $\epsilon_{I2}$. Ignoring the boundary
terms in Eq.~\eqref{GPvar}, the modified GP equation can then be
written in terms of the density $\rho(r)=|\Psi(r)|^2$ as
\begin{equation}
  \label{eq:gpe-dimless}
  {\mu} =\frac{1}{2} r^2+4\pi a \left(\rho+ g_2
    \nabla^2 \rho\right).
\end{equation}
With scaled coordinate $x=r/\sqrt{g_2}$ (assuming $g_2>0$ for the moment) and density $f(r)=4\pi a x\rho(r)/g_2$, this becomes
\begin{equation}
 \frac{\ud^2 f}{\ud x^2}+f= \frac{\mu}{g_2} x -\frac{1}{2} x^3,
\end{equation}
The inhomogeneous and homogeneous solutions with boundary condition
$f(0)=0$ are
\begin{equation}
  \label{eq:f-solution}
  f_i(x)=(\frac{\mu}{g_2}-\frac{1}{2}x^2 +3) x, \quad
  f_h(x)=\frac{A}{g_2} \sin x,
\end{equation}
where $A$ is a constant (with dimensions of length squared) to be
determined later. The full solution is
\begin{equation}
  \label{eq:density-solution}
   \rho( x)=\frac{g_2}{4\pi a}\left[
    \frac{\mu}{g_2} -\frac{1}{2} x^2 +3
    +\frac{A}{g_2}\frac{\sin x}{x}
  \right].
\end{equation}
For a given $A$, the chemical potential $\mu$ and the condensate radius $R$
are determined by the normalization and the surface condition,
\begin{equation}
  \label{eq:norm-surf-cond}
  \int_{0}^{x_0} 4\pi x^2 \rho(x)\ud r=N
  \quad\textrm{and}\quad
  \rho(x_0)=0,
\end{equation}
where $x_0=R/\sqrt{g_2}$.
The solution $\rho$ should be positive for $x<x_0$ which must be
explicitly checked. Outside $x_0$, we use $\rho=0$.

We now consider the boundary terms in Eq.~\eqref{GPvar}. Above, we 
assumed that $\rho(x_0)=0$ at some finite radius $x_0$ which we identify as the
condensate size. However, only the first two boundary terms in
Eq.~\eqref{GPvar} vanish on account of this 
condition.  For the last term in Eq.~\eqref{GPvar} to vanish we
need $\nabla_x\Psi(x_0)=0$, which implies that
\begin{equation}
  \label{eq:surface-derivative}
  \frac{d\rho}{dx}(x_0)=0.
\end{equation}
Notice that this latter derivative is in fact non-zero in the $g_2=0$ case, which is 
the root of the divergence of the kinetic energy at the condensate surface as we discuss later. 
Equation~(\ref{eq:surface-derivative}) gives a closed expression for the remaining free parameter $A$,
\begin{equation}
  \label{eq:A-vs-z}
  \frac{A}{g_2}=\frac{x_{0}^{3}}
  {x_0\cos x_0-\sin x_0}.
\end{equation}

This additional requirement on the derivative at the edge of the
condensate implies that higher-order terms require a smoothing at the
surface of the cloud. 
In addition, the
discussion of which kinetic operator structure to use
($|\nabla\Psi|^2$ or $\Psi^*\nabla^2\Psi$ \cite{lundh97}) is
obsolete in our treatment since the boundary term $\delta\Psi^*\nabla\Psi$
vanishes. In this sense the
inclusion of a higher-order term neatly removes some of the
difficulties of the traditional TF treatment.

The solutions with a finite boundary $R$ of the modified GP equation
only minimize the energy functional if
Eq.~\eqref{eq:surface-derivative} holds. We note that extremal states
of the energy functional always satisfy the virial theorem. Thus,
enforcing the virial theorem on the GP solutions is equivalent to
Eq.~\eqref{eq:surface-derivative}. We show in the Appendix that the virial
theorem approach also leads to Eq.~\eqref{eq:A-vs-z}.

\section{SIZE AND CHEMICAL POTENTIAL}\label{sizechem}
We now determine the condensate size $R$ and chemical potential $\mu$.
The normalization condition is
\begin{equation}
  \label{eq:normalization-condition2}
  \frac{Na}{g_2^{5/2}}=x_0^3\left(\frac{\mu}{3g_2}-\frac{x_0^2}{10}\right),
\end{equation}
while the surface condition reads
\begin{equation}
  \label{eq:surface-condition2}
  \frac{\mu}{g_2} -x_0^2/2+3+\frac{A}{g_2}\frac{\sin x_0}{x_0} =0.
\end{equation}
Combining eq.~\eqref{eq:A-vs-z}-\eqref{eq:surface-condition2} gives
\begin{equation}
  \label{eq:rmax-equation}
  \frac{Na}{g_2^{5/2}}=x_0^3\left(\frac{x_0^2}{15}-1+\frac{x_0^2/3}{1-x_0\cot x_0}\right),
\end{equation}
which determines $R$ for given $Na$ and $g_2$, and upon 
back-substitution also $\mu$.

The $g_2<0$ case can be worked out analogously by replacing trigonometric 
functions with hyperbolics and keeping track of signs. The two cases
can in fact be combined into one equation
\begin{equation}
  \label{eq:rmax-equation2}
  \frac{Na}{|g_2|^{5/2}} 
  =|x_0|x_0^2\left(\frac{x_0^2}{15}-1+\frac{x_0^2/3}{1-|x_0\cot x_0|}\right).
\end{equation}

\begin{figure}[t!]
  \epsfig{file=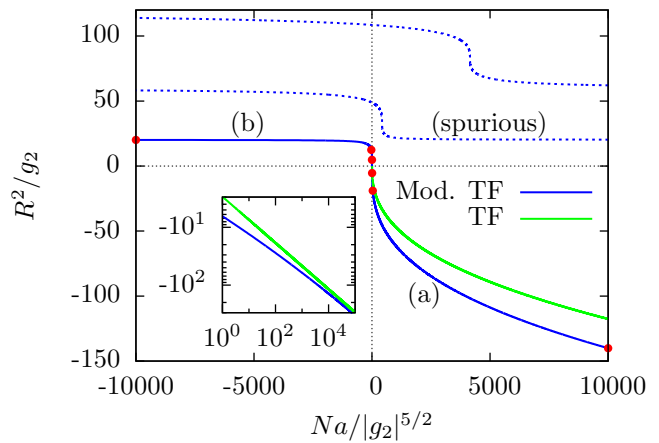,clip=true,scale=1.0}
  \caption{(Color online) 
		Condensate size ($R$) as function of $Na$ and $g_2$ as
    found in the modified TF approximation,
    Eq.~\eqref{eq:rmax-equation2}.  The solutions (a) and (b)
    correspond to the sign combinations ($a>0$, $g_2<0$) and ($a<0$,
    $g_2>0$), respectively.  No solutions exist for $a g_2>0$. The
    spurious solutions (dotted) have negative densities for one or
    more intervals inside $R$. The branch (a) approaches the normal TF
    result Eq.~\eqref{eq:mu-rmax-TF} when $Na\to+\infty$ or $g_2\to
    -0$. Note that the convergence is only relative [see
    eq.~\eqref{eq:rmax-equation2}] and the TF limit is better
    represented in the logarithmic inset. Points indicate the data
    from Tab.~\ref{tab:data}. All values are in trap units.}
  \label{fig-rmax}
\end{figure}

This equation determines $x_0^2=R^2/g_2$ implicitly as function of $N
a/|g_2|^{5/2}$. The result is shown in Fig.~\ref{fig-rmax}. We notice that
in principle, $R$ becomes a multi-valued function. However, all the
higher solutions for $g_2>0$ [dotted in Fig.~\ref{fig-rmax}] are
spurious, since the density becomes negative on one or more intervals
inside $R$. 
The non-spurious solutions [solid line in Fig.~\ref{fig-rmax}] define
$R$ as a single-valued function of $a$ and $g_2$, which was not
guaranteed a priori. 
The four quadrants in Fig.~\ref{fig-rmax} correspond to the different
sign combinations of $a$ and $g_2$. 
The sign of the extra interaction energy, $E_{I2}$, 
is determined by $a
g_2\nabla^2\rho$. For a typical concave density, the Laplacian term will
be negative. We therefore see that for $ag_2>0$, the higher-order 
interaction is attractive, whereas for $ag_2<0$, it is repulsive.
The TF solution only exists for $ag_2<0$. We discuss both cases separately below.

\subsection{The attractive regime: $ag_2>0$}

For $a<0$, $g_2<0$ [third quadrant in Fig.~\ref{fig-rmax}] there are
no solutions, which is expected since the normal TF approximation has
no solutions for $a<0$ as the interaction energy $E_I$ is negative
and the kinetic energy that could prevent collapse is neglected.

The $g_2>0$, $a>0$ case in the first quadrant has only spurious
solutions. Here the $g_2$ term is attractive for the typical concave
density and a collapse towards a high-density state is possible in
complete analogy to the usual discussion of attractively interacting
condensates within the standard GP theory. Whereas there can be
metastable states at large values of $Na/g_{2}^{5/2}$, these are
stabilized by kinetic energy and thus are not present in our TF
approach.  Thus, even when the total kinetic energy is small, it
is still needed to prevent the attractive higher-order term from
amplifying local-density variations.

This important point can also be established by considering the stability of the
homogeneous condensate through linearization of the GP equation. 
By repeating the analysis of \cite{pethick02} with the higher-order term, 
we find that for $g_2>0$ 
and $a>0$, the kinetic-energy term is crucial for the stability 
of the excitation modes. In fact, exponentially growing modes
will always be present if the kinetic energy is neglected.
This will be discussed elsewhere in relation to the
numerical solution of the full GP equation \cite{thoger2009b}.

\subsection{The repulsive regime: $ag_2<0$}
For $g_2<0$, $a>0$ a single solution (a) exists.  This was expected
since $E_{I2}>0$ gives extra stability.  The solution approaches the
normal TF result in Eq.~\eqref{eq:mu-rmax-TF} when
$Na/|g_2|^{5/2}\to+\infty$, as can also be seen from
Eq.~\eqref{eq:rmax-equation2}. Of course in this limit $E_{I2}\ll E_I$. 
However, the $-1$ term in Eq.~\eqref{eq:rmax-equation2} implies that 
the convergence to the normal TF solution
is only on a relative scale and is better represented on a 
logarithmic scale as in the 
inset in Fig.~\ref{fig-rmax}.

For $g_2>0$,
$a<0$ there is a single solution (b) which connects smoothly to the
(a) solution.  In the limit $Na/|g_2|^{5/2}\to -\infty$, which is
determined by $x_0\cot x_0=1$, we find $R^2/g_2=20.1907$. This solution
is possible when the $g_2$ term provides just enough repulsion to
cancel the usual $a<0$ collapse behavior.

\subsection{Chemical potential}
\begin{figure}[t!]
  \epsfig{file=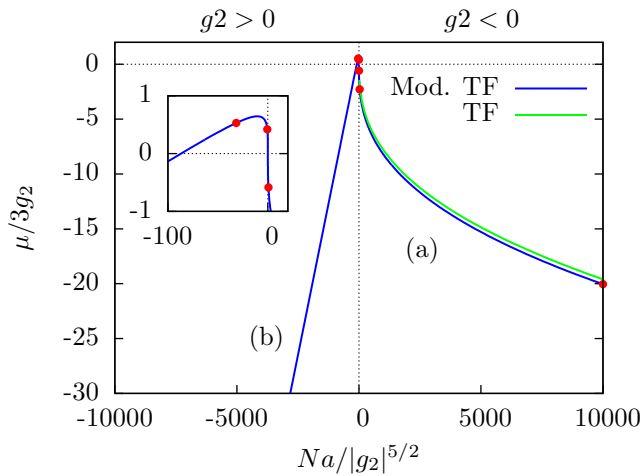,clip=true,scale=1.0}
  \caption{(Color online) Chemical potential $\mu$ as function of $Na$
    and $g_2$ as found in the modified TF approximation, using the
    solutions (a) and (b) from Fig.~\ref{fig-rmax}.  For branch (a)
    and the upper part of branch (b) (see inset), we have $\mu>0$. The
    lower part of (b) has $\mu<0$. Points indicate data from
    Tab.~\ref{tab:data}. All values are in trap units.}
  \label{fig-mu}
\end{figure}

In Fig.~\ref{fig-mu} we show the chemical potential for the smoothly
connecting solutions (a) and (b). Again we see that (a) approaches the
normal TF limit for large $Na/|g_{2}|^{5/2}$. Here, it is
interesting to note how $\mu$ turns around near the origin
[amplified in the inset in Fig.~\ref{fig-mu}] and maintains a positive 
value. This occurs in the
region where the lowest-order interaction gives a large
negative-energy contribution which the $g_2$ term is still able 
to balance yielding a well-defined TF solution. This behavior is 
analogous to the balancing of attraction by the kinetic term 
in the usual $a<0,g_2=0$ case \cite{baym96,dalfovo99}. As $a$ 
becomes increasingly negative, so too does $\mu$ and collapse
is inevitable (and likewise when $g_2\rightarrow 0^+$).

\section{DENSITIES AND ENERGIES}\label{densene}
With $R$ and $\mu$ determined, we can find the density profile, energy
densities and integrated energy contributions. With 
Eq.~\eqref{eq:density-solution}, the energy densities are given by
\begin{eqnarray}
  &\epsilon_V=\frac{ x^2}{2} \rho,\qquad
  \epsilon_I=2\pi  a \rho^2,\\
  &\epsilon_{I2}  =-\frac{1}{2} \rho(3
  +\frac{A}{ g_2 } \frac{\sin x}{x}).
\end{eqnarray}
Using Eq.~\eqref{eq:gpe-dimless}, the total energy density (without
$\epsilon_K$) becomes
\begin{equation}
  \label{eq:energy-dens}
  \epsilon \equiv  \epsilon_V+\epsilon_I+\epsilon_{I2}
  =\frac{1}{2} \rho(x) (V(x)+\frac{\mu}{g_2}).
\end{equation}

\begin{figure}[t!]
  \epsfig{file=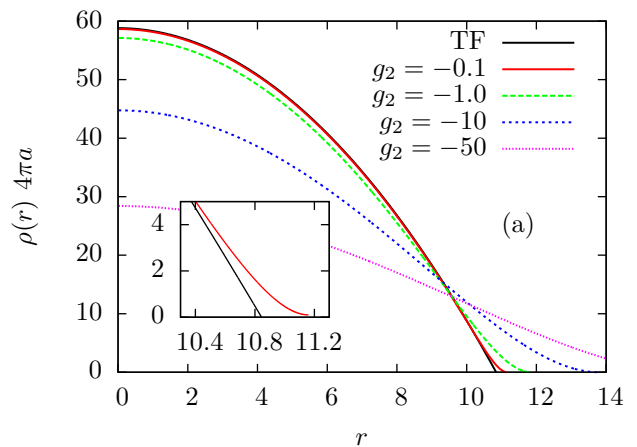,clip=true,scale=1.0}
  \caption{(Color online) Densities for branch (a) in
    Fig.~\ref{fig-rmax} ($g_2<0$ and $Na=10^4$). The $g_2=-0.1$ curve
    is on top of the normal TF result. Inset shows the smooth behavior at the surface
    for $g_2<0$.
    All values are in trap units.}
  \label{fig-dens-a}
\end{figure}

In Fig.~\ref{fig-dens-a}, we show the density profile of the (a) solutions 
for $Na=10^4$ and selected $g_2<0$. We clearly see that the higher-order
term tends to expand the condensate through its repulsion. Importantly,
at the boundary, there is a smoothing caused by the 
condition in Eq.~\eqref{eq:surface-derivative} 
[see inset in Fig.~\ref{fig-dens-a}]. We will
discuss how this affects the estimated kinetic energy in the next section.
As $|g_2|$ grows, we see the condensate flatten and in the limit of 
very large $|g_2|$, it becomes a constant density. 

\begin{figure}[t!]
  \epsfig{file=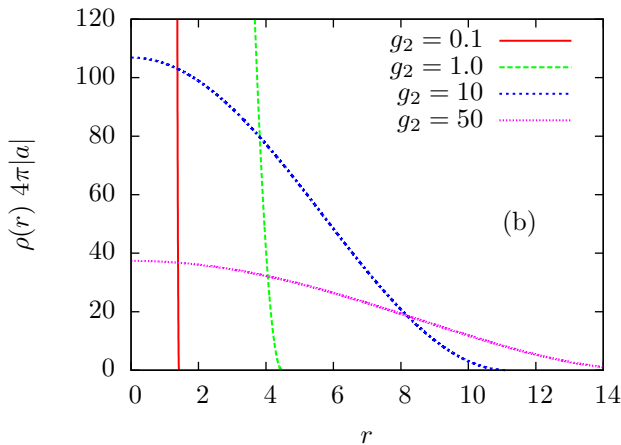,clip=true,scale=1.0}
  \caption{(Color online) Same as Fig.~\ref{fig-rmax} but for
    solutions (b), i.e., opposite signs $g_2<0$ and $Na=-10^4$. }
  \label{fig-dens-b}
\end{figure}

Figure~\ref{fig-dens-b} displays the density profile for the (b) solutions with
$a<0$ for selected $g_2>0$. Here, we see the profile collapse toward the 
expected delta-function with decreasing $g_2$. 
It is interesting to follow the (a) solution through the origin in Fig.~\ref{fig-rmax} and 
onto solution branch (b), passing from
$g_2=-\infty$ to $g_2=\infty$. On the (a) branch, the solution flattens as $g_2$ decreases
and eventually becomes effectively constant in space. This is also true for the (b) branch 
at $g_2=\infty$, and as $g_2$ is decreased, the solution proceeds to shrink as the 
$g_2$ term becomes unable to provide the repulsion needed to prevent the $a<0$ collapse
induced by the lowest-order term.

From the figures, we see that large $|g_2|$ induces large changes in cloud size. As the condensate
can be imaged with very good resolution \cite{hau98}, this should be measurable if the 
regime of large $|g_2|$ can be accessed.

{\renewcommand{\tabcolsep}{0.3cm}
\begin{table*}[t!]
  \centering
  \begin{tabular}{cc|cccccccc}
    \hline \hline
   &$g_{2}$&$R$&$\mu$&$E_{V}/N$&$E_{I}/N$&$E_{I2}/N$&$E/N$&$E_R/N$&$E_{K}/|E|$\\
    \hline
    TF&---&10.8447&58.8040&25.2017&16.8011&---&42.0028&16.8012&3.135$\times10^{-3}$ \footnotemark[1] \\
    \hline 
       &$-0.01$               &10.9447&58.8188&25.2164&16.7865&0.01465&42.0176&16.8012& 1.8$\times10^{-3}$\\
       &$-0.1$                &11.1607&58.9481&25.3430&16.6635&0.13909&42.1456&16.8026& 1.4$\times10^{-3}$\\
    (a)&$-1.0$ \footnotemark[2]&11.8364&60.1210&26.4309&15.6818&1.16330&43.2760&16.8451& 1.0$\times10^{-3}$\\
       &$-10$ \footnotemark[2] &13.7835&68.4515&32.9856&11.3469&6.38609&50.7186&17.7330& 0.57$\times10^{-3}$\\
       &$-50$ \footnotemark[2] &16.439 &87.8248&45.6836&6.99293&14.0777&66.7542&21.0706&0.30$\times10^{-3}$\\
    \hline
       &50 \footnotemark[2]  &15.407 & 63.0102&38.9723 &$-8.92496$&20.9439&50.9912 &12.0189&0.43$\times10^{-3}$\\
       &10 \footnotemark[2]  &11.170& 15.9128&19.8375 &$-24.7434$&22.7810&17.8751 &$-1.9623$&2.3$\times10^{-3}$\\
    (b)&5.14 \footnotemark[3]&9.1999 &$-13.1384$&13.1579 &$-46.0283$&32.8801&0.0097  &$-13.148$&6.098\\
       &1.0 \footnotemark[2] &4.4801 &$-327.612$&3.04199 &$-416.359$&251.032&$-162.285$&$-165.33$&1.5$\times10^{-3}$\\
       &0.1                 &1.4204 &$-10456.4$&0.30571&$-13071.1$&7842.81&$-5227.98$ &$-5228.4$&0.47$\times10^{-3}$\\
    \hline \hline
  \end{tabular}
  \footnotetext[1]{The kinetic energy estimated by surface corrections as in \cite{pethick02}.}
  \footnotetext[2]{Values are indicated by points in Figs.~\ref{fig-rmax} and \ref{fig-mu}.}
  \footnotetext[3]{The total energy $|E|$ is zero near $g_2=5.14$, hence the TF approximation is
    invalid here.} 
  \caption{Chemical potential $\mu$ and condensate size $R$
    for different $g_2$ and fixed
    $N|a|=10^4$. 
		The integrated energies are trap ($E_V$), interaction ($E_I,E_{I2}$), total ($E=E_V+E_I+E_{I2}$), and
		release energy ($E_R=E-E_V$). 
    The TF limit is approached for $g_2\to -0$. The ratio
    of kinetic energy $E_K$ to total energy $E$ indicates
    where the TF approximation is valid. The corresponding density
    distributions are shown in Figs.~\ref{fig-dens-a} and
    \ref{fig-dens-b}. All values are in trap units.}
  \label{tab:data}
\end{table*}
}

\begin{figure}[t!]
  \epsfig{file=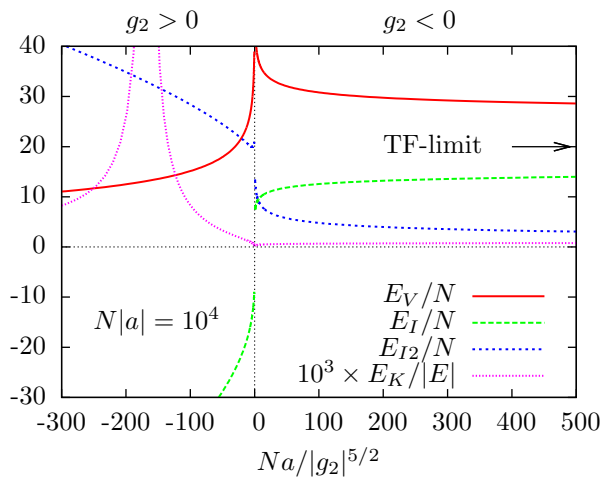,clip=true,scale=1.0}
  \caption{(Color online) Different total-energy contributions.  $N|a|=10^4$. All values are in trap units.}
  \label{fig-energy}
\end{figure}

We now discuss the energy contributions which are interesting since the 
release energies are in fact measurable quantities \cite{pitaevskii03}. Since we neglect
the kinetic term in the TF approximation, the release energy is simply $E_R=E_I+E_{I2}=E-E_V$.
In Tab.~\ref{tab:data}, we give the integrated energy contributions for some relevant values of $g_2$
calculated for $N|a|=10^4$, whereas Fig.~\ref{fig-energy} gives the energies as function of $Na/|g_2|^{5/2}$.
We note that for smaller values of $N|a|$, the same overall behavior is found,
however, the kinetic term is more important and the TF approximation becomes worse.

We observe that $E/N$ grows towards the $|g_2|=\infty$ point. This is due to the trap energy increasing
as the density flattens [$E_V$ diverges around the origin in Fig.~\ref{fig-energy}]. 
Furthermore, as $g_2\rightarrow 0^+$ the energy diverges toward $-\infty$
as the collapse sets in [$E_I$ diverges on the $g_2>0$ side in Fig.~\ref{fig-energy}]. 
The boundary where the energy vanishes is around $g_2=5.14$ for $N|a|=10^4$, but
this depends on the choice of $N|a|$. With respect to the release energy, we find that somewhere in 
the region $10<g_2<50$, $E_R$ becomes negative. This is a result of the unavoidable collapse, and
also indicates that kinetic energy cannot be ignored at this point. Notice, however, that 
the release energy changes considerably and could provide a way to measure the influence of the 
$g_2$ term.

\section{CONSISTENCY OF THE THOMAS-FERMI APPROXIMATION}\label{cons}
We now address the validity of the TF approximation with the $g_2$ term included. In 
order to do so, we must consider the contribution of the kinetic energy.
The kinetic energy density can be written as
\begin{equation}
  \label{eq:eK}
  \epsilon_K
  =\frac{g_2}{8 \rho(4\pi a)^2}\left(
     x+\frac{A}{g_2}
    \frac{x
    \sin x
    -\cos x}{x}
  \right)^2.
\end{equation}
Strictly speaking, this is not the true kinetic energy, since the
kinetic terms were neglected from the start. However
Eqs.~\eqref{eq:energy-dens} and \eqref{eq:eK} can be used to test
whether the TF approximation holds locally, i.e., $\epsilon_K \ll
\epsilon$ should hold for the solution $\rho$ to be consistent. In Tab.~\ref{tab:data},
we calculate the integrated contribution of the kinetic energy relative to 
the total TF energy and we find that the contribution is small everywhere except 
the point where $E=0$ on the $g_2>0$ side of Fig.~\ref{fig-energy}. Here, the kinetic energy is of course 
the most important term and the TF approximation is poor.

In the standard TF, the kinetic energy causes trouble at the boundary of the cloud.
Here, $\nabla\Psi\propto \nabla\rho/\sqrt{\rho}$ and since the density vanishes 
and the derivative is finite [see Eq.~\ref{eq:density-TF}], this diverges at $R_{TF}$.
When including the higher-order term we need to use the additional boundary 
condition $\nabla\Psi=0$ at $R$, so the kinetic energy will be strictly zero
at $R$. However, as one approaches the boundary, the kinetic-energy density grows 
rapidly before it descends towards zero within a very small interval 
at $R$. The total energy density in Eq.~(\ref{eq:energy-dens}) goes to zero at this point 
and we find that $\epsilon_K/\epsilon$ is very large near the boundary as in 
the usual $g_2=0$ case.

We conclude that the inclusion of the higher-order term does not alleviate the 
difficulties with kinetic energy at the boundary. The techniques for 
addressing this problem described in \cite{dalfovo96,lundh97} should therefore
be generalized to include the higher-order interaction term in order to 
improve the description at the boundary of the cloud.

\section{COMPARISON TO ATOMIC SYSTEMS}\label{compare}
The considerations above show that deviations from the usual TF approximation
can be strong when $g_2$ is large. In the following, we reintroduce explicit units
for comparison with real systems. We have to consider $g_2/b^2$.
Of course, the $b^2$ factor means
that this quantity is generally very small since $g_2$ is of order $a_{0}^{2}$
and $b$ is of order $10^4 a_0$.

We first consider some typical 
background values for bosonic alkali atoms away from resonance. We estimate
the effective range to be the of order of the potential range and assuming a 
van der Waals interaction, we have $r_e\sim 50-200a_0$.
For typical one-component gases we have 
$-450a_0\lesssim a \lesssim 2500a_0$ \cite{chin2009}. Since $g_2=a^2/3-ar_e/2$,
we see that the $a^2$ term will dominate and in all cases
$0<g_2\lesssim 10^6a_0$. In trap units, this becomes
$g_2/b^2\lesssim 5\cdot10^{-3}(1\mu\text{m}/b)^2$.
In typical traps of $b\sim 1-10\mu\text{m}$, 
the higher-order term is therefore very small. 
These values also predominantly lie in the first quadrant of 
Fig.~\ref{fig-rmax} and thus no TF solution exists. 

Let us first consider Feshbach resonances in order to increase the influence of
the $g_2$ term.
We use a multi-channel Feshbach model \cite{bruun05},
which describes the full T matrix 
as a function of resonance position
$B_0$, width $\Delta B$, magnetic-moment difference between the
channels $\Delta \mu$, and the background scattering length $a_{bg}$.
Performing an effective-range expansion \cite{zinner2009},
we have $a=a_{bg}[1-\Delta B/(B-B_0)]$ and
$r_e=r_{e0}/[1-(B-B_0)/\Delta B]^2$, where
$r_{e0}=-2\hbar^2/ma_{bg}\Delta \mu\Delta B<0$. 
Combining these relations, we find $r_e=r_{e0}(1-a_{bg}/a)^2$ and
\begin{equation}
  \label{eq:g2(a)}
  g_2(a)=\frac{a^2}{3}-\frac{a r_{e0}}{2}(1-\frac{a_{bg}}{a})^2.
\end{equation}
Hence $g_2$ diverges when $a\to 0$ (referred to as zero-crossing) or
$a\to\infty$ (on resonance). Near zero crossing, the effective-range
expansion is, however, severely divergent and its validity is questionable.
Even so, the effective-range corrections near zero-crossing obtained
are in fact identical to those obtained from use of the full T-matrix \cite{zinner2009b}.
One finds $\lim_{a\rightarrow0}a g_2=|r_{e0}|a_{bg}^2/2$, where $r_{e0}<0$.

As a concrete example, we consider
the alkali isotope $^{39}$K where several Feshbach 
resonances of vastly different widths were found recently \cite{errico07}.
First, we focus on zero-crossing and consider the very narrow resonance at $B_0=28.85$G with 
$\Delta B=-0.47$G, $\Delta \mu=1.5\mu_B$, and $a_{bg}=-33a_0$.
We obtain
$r_{e0}=-5687a_0$ and $ag_2\rightarrow 93.8\cdot10^3a_{0}^{3}$ for $a\rightarrow 0$.
It is important to notice that $ag_2>0$ around $a=0$. This means that we are looking
for solutions in the first and third quadrants of Fig.~\ref{fig-rmax} and again we
have to conclude that no TF solutions can be found when higher-order terms are
taken into account.

Another case of interest is around resonance where $|a|=\infty$. Here, we have
$r_e\sim r_{e0}$ and $g_2\propto a^2>0$ on both sides of the resonance. Thus, the
$a>0$ side will be in the first and the $a<0$ in the second quadrant of Fig.~\ref{fig-rmax}.
This makes it difficult to imagine sweeping the resonance from either side to 
probe the solutions on branch (b) in Fig.~\ref{fig-rmax}. One could imagine starting on the $a>0$ 
side with small $g_2>0$. The full GP equation will have perfectly sensible solution here, however,
when one approaches the resonance the $g_2$ term will diverge and induce collapse 
already on the $a>0$ side. If we approach from the $a<0$ side, then we face the problem
that the critical number of particles decreases dramatically before $g_2$ grows
sufficiently and one therefore needs a very small condensate since
$Na/b\sim 0.5$ \cite{zinner2009}. At this point, the TF approximation is no longer valid.

The Feshbach resonance used to increase $g_2$ must be very narrow in order for $r_{e0}$
to be large. However, most experimentally known resonances are not narrow. For 
broad or intermediate resonances, we have to consider the long-range van der Waals interaction
when calculating the effective-range corrections. Analytic formulas for this case have been
worked out in \cite{gao98}, and we note that the effective range diverges as $a^{-2}$ near zero crossing 
exactly as
in the Feshbach model above. For very narrow resonances, we still have $\beta_6\ll r_{e0}$, where
$\beta_6$ is the characteristic length of the van der Waals interaction.
The model above should thus give the dominant contribution.

Using the van der Waals formulas we can estimate $ag_2$ at zero crossing. We find
\begin{align}
\frac{ag_2}{b^3}\rightarrow -\frac{1}{3x_e}\left(\frac{\beta_{6}}{b}\right)^3,
\end{align}
where $x_e=(\Gamma[1/4])^2/2\pi$, with $\Gamma$ the gamma function.
We have explicitly introduced the oscillator length which is the relevant length scale 
of comparison. Importantly, we find that $ag_2<0$ for $a>0$ and we are thus in the fourth 
quadrant where a TF solution exists. 
For $a<0$, we pass to the second quadrant as $g_2$ becomes positive and
a single collapsed solution can be found.

We now estimate the parameters obtained from the van der Waals formulas. 
With $b=1\mu$m and $\beta_{6}\sim 123.3a_0$ \cite{pethick02},
we have $ag_2/b^3\sim -10^{-8}$. We thus have $Na/|g_2|^{5/2}\propto 10^{8} (Na) a^{5/2}$. 
For values of $a$ that are not extremely small, the solution is therefore
typically located far to the right in Fig.~\ref{fig-rmax} where it will look similar
to the $g_2=0$ case.
We can estimate how close to 
zero one would have to tune $a$ in order to see deviations using the $a\rightarrow 0$ limit
of the van der Waals effective range. Let us aim for $g_2/b^2=- 10$ which should be observable
in the condensate profile according to Fig.~\ref{fig-dens-a}. With $b=1\mu$m, we need 
$a\sim 10^{-6}\beta_6\sim 1.7\times 10^{-4}a_0$. Using broad resonances, one can tune 
to zero at the level of $10^{-2}a_0$ in $^{39}$K \cite{fattori2008}. Observing the effect of the 
$g_2$ term therefore seems out of reach at the moment, but might be possible in the near future.
Of course, we still have to maintain a large value of $Na$ for kinetic energy to be small, 
and thus a larger condensate is needed close to zero crossing.

From the examples above, we see problems in accessing the TF solutions presented
above in current experiments with ultracold alkali gases. 
In particular, we notice that
realistic systems which have been used for creation of BECs in alkali-metal gases for the last
decades have parameters that predominantly lie 
in the first quadrant of Fig.~\ref{fig-rmax}. As we have discussed, there are no well-defined TF
solutions in that region. Therefore, we see that the kinetic energy plays 
a decisive role and we are forced to consider it in principle, even if it is small 
for all practical purposes. The physical reason is that for $a>0$ and $g_2>0$, the 
higher-order interaction is effectively attractive and induces collapse which will have to  
be balanced by a barrier from the kinetic term, similar to the $a<0$, $g_2=0$ case
\cite{dalfovo99}. Since we neglect the kinetic term in the TF approximation, we should not
expect to find solutions in the $ag_2>0$ case.

Only in the case of resonances 
dominated by the long-range van der Waals interaction do the parameters allow 
for TF solutions with non-zero $g_2$. However, here the length scale of the 
trap makes the contribution very small and the TF solution becomes identical
to the $g_2$ case. One could in principle tune $a$ very close to zero-crossing
and obtain a significant contribution but the level of tuning required is 
beyond current experimental reach.

\section{CONCLUSIONS}\label{conc}
We have considered the effect of higher-order interactions in Bose-Einstein 
condensates within the Gross-Pitaevskii theory. We derived the GP
equation with effective-range corrections included and solved it 
analytically in the Thomas-Fermi approximation. 
Higher-order interaction terms act as
derivatives on the condensate wave function which means that the boundary conditions
on the solutions of the GP equation must be carefully considered. 
We then discussed the 
solutions for various parameters and presented the chemical potential, density profiles,
and the energy contributions. 

We find that no TF solutions 
are possible when the higher-order term is attractive. This conclusion holds 
both in the
trapped system and in the homogeneous case \cite{thoger2009b}.
An estimate of the 
relevant parameters for alkali atoms showed that away from resonances,
they typically lie in the region where the effective-range correction
is effectively attractive and likewise near very narrow Feshbach 
resonances. We conclude that in those cases,
the kinetic energy, even if very small, is crucial in 
order to stabilize collapse due to higher-order interaction terms. 
For broader resonances where the long-range van der Waals
potential is dominant, we find that modified TF solutions exits. However, 
for typical traps, the parameters are very small and tuning of the
scattering length near zero crossing at a level beyond current experimental 
reach is necessary. This might of course become possible as experimental 
control improves in the future.

\paragraph*{ACKNOWLEDGMENTS}
Discussions with D. V. Fedorov, N. Nygaard, and  I. Zapata are highly
appreciated.

\appendix

\section{DETERMINATION OF $A$ FROM THE VIRIAL EQUATION}
Even though Eq.~\eqref{eq:density-solution} is a solution to the
modified GP equation Eq.~\eqref{eq:gpe-dimless} for all $A$, it does
not necessarily minimize the energy functional as discussed in
the main text. This can also be seen from
the virial equation (with neglected kinetic energy),
\begin{eqnarray}
  \label{eq:mgp-virial}
  -2E_V+3E_I+5E_{I2}=0,
\end{eqnarray}
which holds for all extremal points of the energy
functional. Equation~\eqref{eq:mgp-virial} is derived from the energy
functional using scaling arguments as in \cite{pitaevskii03}.

As an example, consider the $A=0$ solution in
Eq.~\eqref{eq:density-solution}. This solution has a chemical
potential shifted by $3 g_2$ compared to the $g_2=0$ TF result. But
the density is unchanged and so is $E_V$ and $E_I$. Hence, the usual
virial equation $-2E_V+3E_I=0$ for $ g_2=0$ also holds for $ g_2\ne
0$. Since $E_{I2}=-3 g_2/2 \ne 0$, the virial equation
Eq.~\eqref{eq:mgp-virial} is not fulfilled, and hence the $A=0$
solution is not extremal. Below, we use the virial equation to calculate 
the value of $A$ that minimizes the energy functional and the 
corresponding $R$ and $\mu$. We will also prove that this condition
is in fact equivalent to assume $\rho(x_0)=\nabla_x\rho(x_0)=0$ at the 
boundary.

The general results for $A$, $R$, and $\mu$ can  be derived using the
normalization and surface conditions Eq.~\eqref{eq:norm-surf-cond}, and
the virial equation Eq.~\eqref{eq:mgp-virial}.
For convenience, we introduce the variables
$\bar\mu = \mu/(3 g_2)+1$, $ \bar A =A/( 3g_2)$, and
$c=N a/g_2^{5/2}$. 
The different energy contributions are
\begin{equation}
  \label{eq:energy-integrals}
  \begin{split}
    E_V&=3s\int_0^{x_0} \ud x x^4(\bar\mu -\frac{x^2}{6}
    +\bar A \frac{\sin x}{x}),\\
    E_I&=9s\int_0^{x_0} \ud x x^2(\bar\mu -\frac{x^2}{6}
    +\bar A \frac{\sin x}{x})^2,\\
    E_{I2}&=-9s\int_0^{x_0} \ud x
    x^2(\bar\mu -\frac{x^2}{6}+\bar A \frac{\sin x}{x})
    (1+\bar A \frac{\sin x}{x}),
  \end{split}
\end{equation}
where $s= g_2^{7/2}/(2 a)$. Direct integration of
Eq.~\eqref{eq:energy-integrals}, insertion of $\bar\mu$ from
Eq.~\eqref{eq:surface-condition2}, and some algebra gives the virial
equation
\begin{equation}
  \label{eq:Evir}
  \begin{split}
    0&= -2E_V+3E_I+5E_{I2}\\
    &=-\frac{s}{x_0}(x_{0}^{3}-3\bar A (x_0\cos x_0-\sin x_0))^2.
  \end{split}
\end{equation}

We immediately see that this is in fact 
equivalent to Eq.~\eqref{eq:A-vs-z}. Therefore, the solution 
we have explicitly found above minimizes the energy functional with 
boundary conditions $\rho(x_0)=\nabla_x\rho(x_0)=0$. More generally, when
we solved the modified GP equation without considering the boundary terms in Sec.~(\ref{TFA}), we found a one-parameter
family of solutions (parametrized by $A$). The virial theorem is merely
a constraint on $A$ for obtaining a minimum of $E$.

\end{document}